\documentclass[journal=amlccd,manuscript=letter,layout=onecolumn]{achemso}
\usepackage[version=3]{mhchem}
\usepackage{graphicx}
\usepackage{multicol}
\usepackage{amssymb,amsmath}
\usepackage{xcolor}
\usepackage{soul}
\setkeys{acs}{articletitle=true}

\renewcommand{\hl}[1]{#1}

\title{Nonlinear mechanics of triblock copolymer elastomers: from molecular simulations to network models}

\author{Amanda J.~Parker}
\email{parkeram@physics.ubc.ca}
\affiliation[UBC]
{Dept. of Physics and Astronomy, University of British Columbia, 6224 Agricultural Road, Vancouver BC V6T 1Z1, Canada}
\alsoaffiliation[QMI]
{Quantum Matter Institute, University of British Columbia, Vancouver, BC V6T 1Z4, Canada}
\author{J\"org Rottler}
\affiliation[UBC]
{Dept. of Physics and Astronomy, University of British Columbia, 6224 Agricultural Road, Vancouver BC V6T 1Z1, Canada}
\alsoaffiliation[QMI]
{Quantum Matter Institute, University of British Columbia, Vancouver, BC V6T 1Z4, Canada}

\begin{document}

\abstract{We introduce an entropic network model for copolymer elastomers based on the evolution of microscopic chain conformations during deformation. We show that the stress results from additive contributions due to chain stretch at the global as well as entanglement level. When these parameters are computed with molecular simulations, the theory quantitatively predicts the macroscopic stress response. The model requires only one elastic modulus to describe both physically crosslinked triblock networks and uncrosslinked homopolymers. }

\newpage
Entropic network models of elastomers attempt, with varying degrees of complexity, to connect mechanical response to the behavior of polymer chains during deformation. However, the microscopic chain configurations cannot easily be measured experimentally. Therefore, theoretical models are typically fit to at least the two elastic moduli $G_e$ and $G_c$ relating to the stress contributions of entanglements and crosslinks. The recent non-affine strain model \cite{dav13} fits to experimental data for vulcanized rubber well into the large strain regime. It combines the non-affine tube \cite{rub02} and the Arruda-Boyce 8-chain \cite{arr93} models  so that the former becomes applicable for larger strains. The 8-chain model takes into account the effect of finite chain length, but only considers crosslinks, while the non-affine tube model accounts for the free energy cost of an entangled chain confined to a deforming tube.    

Molecular dynamics simulations of coarse-grained polymers reproduce the experimental stress response of melts \cite{gre90, due91, due94}, copolymers \cite{gre96} and elastomers \cite{cha14,par15,li16} qualitatively,  while also providing insight into the deformation at chain level. However, even here, fitting to a model is required to separate the stress contributions from crosslinks and entanglements between chains \cite{cha14}. In addition, while crosslinks are easy to identify and track during a molecular dynamics simulation, entanglements are not. Methods based on primitive path analysis (PPA) \cite{eve04} that rely on Gaussian chain statistics no longer apply during significant chain deformation. Progress has been made recently utilizing Kr\"oger's Z1-method \cite{kro05,sha07,hoy09,kar09} to identify changes to the entanglement length and tube diameter during deformation \cite{cha14,li16,dav16}. 

\begin{figure}[t]
\begin{center}
\includegraphics[width=0.5\linewidth]{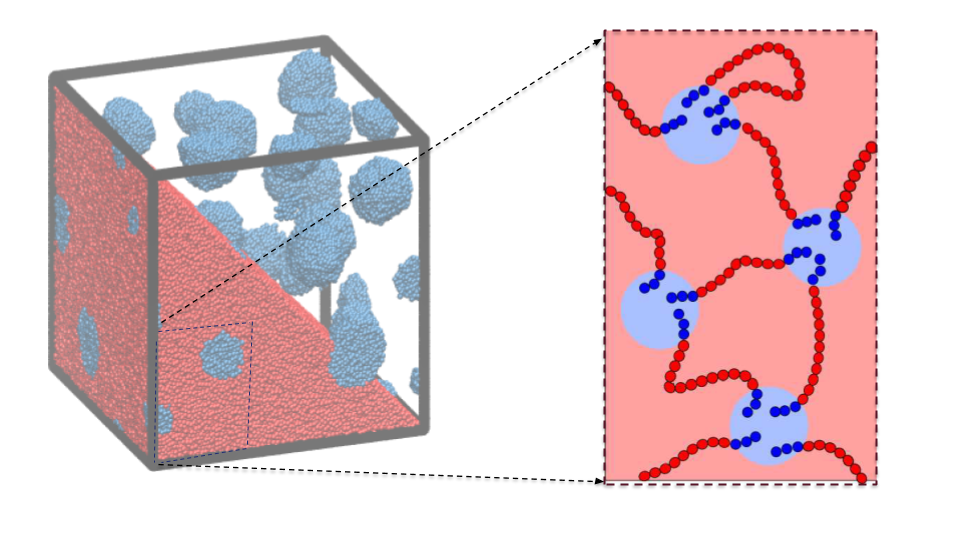}
\caption{ \label{cartoon}
Equilibrated initial configuration of 600 ABA triblock chains with N=800 beads each forming a spherical morphology.
Red: majority soft monomers (90\%). Blue: minority glassy monomers (10\%).}
\end{center}
\end{figure}

In this Letter, we introduce an entropic network description of the stress response of triblock copolymers to volume-conserving uniaxial strain in terms of the change in separation of chain ends and entanglement points.  We use molecular dynamics simulations to track both of these parameters throughout the deformation, and use the microscopic chain level deformation as input into the macroscopic constitutive law. Our description requires only one elastic modulus to describe the contribution of both entanglements and crosslinks to the stress. 

We focus on the common ABA triblock copolymer elastomer, \hl{where a minority phase of $\sim$10-20\% styrenic end-blocks aggregates into spheres embedded in a matrix formed by the rubbery midblock}, see Figure \ref{cartoon}. 
The glassy styrenic regions act as physical crosslinks. Early experimental stress-strain curves were fit to empirical models \cite{hol69,che77}.  More recently, the slip-tube model developed for vulcanized rubbers has been applied to describe both experimental and simulation results for triblock copolymer elastomers. It was found to be a good description for uniaxial deformation of SIS triblock polymers for intermediate stretch ratios ($\lambda = 2.25-4$) \cite{roo05}. Ch

\begin{figure}
\begin{center}
\includegraphics[width=0.5\linewidth]{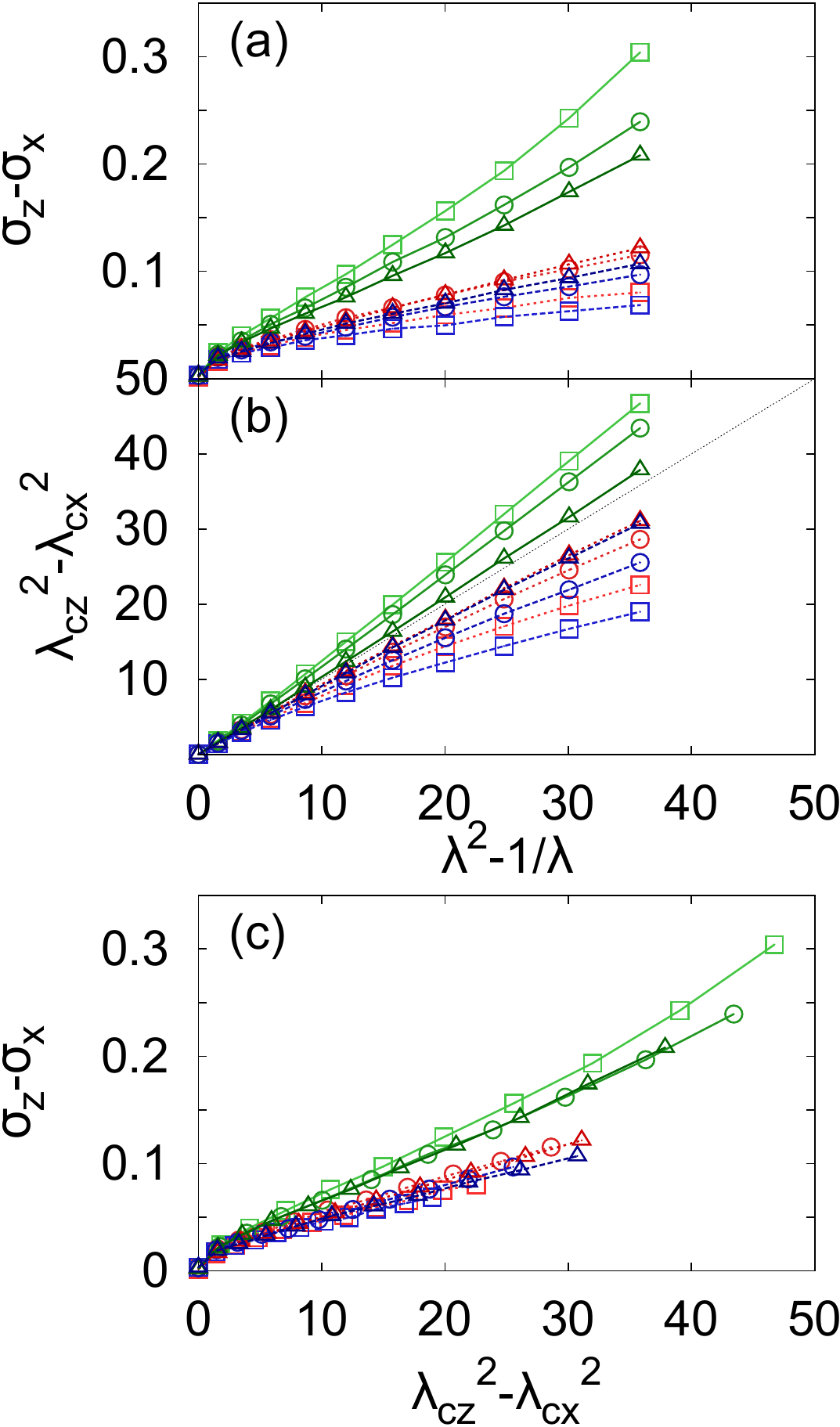}
\caption{
\label{stress}
(a) Stress response in terms of the entropic elasticity factor $g(\lambda)=\lambda^2-1/\lambda$. (b) Effective stretch $g_{\textrm{eff}}=\lambda_{cz}^2-\lambda_{cx}^2$ in terms of components of the chain end-to-end stretch vs. $g(\lambda)$, black line $g(\lambda)= g_{\textrm{eff}}$. (c) Stress vs. effective stretch. Green: triblocks, red: homopolymers, blue: cut chains for $N=300$ ($\square$), $N=500$ ($\circ$), and $N=800$ ($\triangle$), \hl{lines are a guide to the eye.}}
\end{center}
\end{figure}

ABA triblocks are modeled with the standard Kremer-Grest bead-spring model with a Lennard-Jones (LJ) pair potential acting between beads, and a FENE-spring potential acting along the backbone of the chains \cite{auh03}. The LJ potential is truncated at $r_0$=1.5 to include an attractive regime, and all results are quoted in reduced LJ units. Values of the energy parameter were set to $\epsilon_{AA}=1.0$ and $\epsilon_{BB}=0.5$  to give a glass transition temperature that differs by a factor of two. This roughly approximates the ratio of glass transition temperatures in polystyrene and polybutadiene, and $\epsilon_{AB}=0.2$ drives the phase separation. Deformations are performed at $T=0.29$ in between the glass transition temperatures of the hard and soft phases.  

We consider three triblock chain lengths each having 10\% glassy monomers: 15-270-15, 25-450-25, 40-720-40. Each simulation box contains 480,000 monomers. All chains are longer than the entanglement length, $7-18\,N_e$, \hl{which is found using the Z1 method (see below)}. We compare to homopolymers $N=300, 500, 800$ made up of the majority monomers, and a `cut` system which has the same morphology as the triblocks, but chains are cut into separate A,B,A parts after the equilibration. To obtain initial configurations, the chain conformations and phase-separated regions are equilibrated with HOOMD \cite{and08,gla15} using a soft potential following the method described in ref. \cite{par14} before the above model with hard excluded volume interactions is introduced. The number density after equilibration is ramped from an initial value of $\rho=0.85$ to 1.0 to ensure a positive pressure throughout the deformations. Following this, the temperature is quenched from $T=1.0$ to 0.29 at a rate of $10^{-4}$. 

We then use LAMMPS \cite{bro11,pli95} to apply a volume conserving uniaxial strain at an engineering strain rate of $10^{-5}$ (results for $10^{-4}$ are qualitatively identical). The global stretch varies as $\lambda=\lambda_z=1/\sqrt{\lambda_x}=1/\sqrt{\lambda_y}$. Figure \ref{stress}(a) gives the stress response of the material to this deformation. In terms of $g(\lambda)=\lambda^2 - \lambda^{-1}$ the stress is linear for the longest homopolymers when $g(\lambda)>3.5$.
However, the slope of this curve decreases with decreasing chain length, which would indicate a changing value of the elastic modulus $G$ which we expect to be a fixed material property independent of chain length. The cut chains show a very similar response to the homopolymers, markedly different from that of the triblocks, where the hardening is more pronounced. Again the response is mostly linear for the longest chains, and the hardening varies with chain length. Here the shortest chains ($N=300$) exhibit the strongest hardening response. 

Given the chain length dependence of the stress we consider the change in chain end-to-end vectors $\mathbf{R_c}$. We compute an effective stretch $g_{\textrm{eff}}=\lambda_{cz}^2 - \lambda_{cx}^2 $, where $\lambda_{cz}$ and $\lambda_{cx}$ are the rms component-wise, \hl{ensemble-averaged} stretches of the chain \cite{hoy07}. This is calculated for the relevant soft part of the network: the full chain for homopolymers, the mid-blocks of the triblocks and the B-type chains in the cut systems. Figure \ref{stress}(b) shows how the effective chain stretch varies in relation to the global stretch. All homopolymers and cut chains display sub-affine stretch throughout the deformation but the longest chains approach the affine limit.  The triblock chains, by contrast, display super-affine behavior for the shorter chains, which is due to the strong anchoring of the midblock between the almost rigid glassy spheres. 
The relationship between the deformation of glassy regions and the degree of affinity of chain stretch has been explored in detail in ref.~\cite{par15}. Here we focus on the trends with increasing chain length, towards affine in all cases.  

Figure \ref{stress}(c) shows the stress response in terms of the effective stretch \hl{$g_{\textrm{eff}}$} rather than the global stretch. The data collapses onto two curves: one for triblocks and one for both the cut and homopolymer chains. The presence of the glassy spheres is therefore not the dominant cause of the difference in the stress response of the triblocks compared to the homopolymers. Some models include a correction for the styrenic end-blocks acting as a inert filler \cite{hol69, roo05}. They invoke the Guth-Smallwood equation, which increases the elastic modulus by a factor $(1+2.5\phi +14.1\phi^2)$, where $\phi$ is the volume fraction of end-blocks. Since we find no difference between cut and homopolymer chains, this correction is unnecessary.  We now focus on the polymer network deformation to explain why the triblock and homopolymer/cut-chain responses differ.       
     
Accounting for any non-affinity in the deformation of chain ends significantly decreases the chain length dependence of the triblock stress response and removes it for the homopolymers. The homopolymer stress-strain relationship remains linear for all data collected and all chain lengths, but the triblocks still exhibit a stronger nonlinear hardening. Though all chains are now described as having a linear stress response at least over some range of stretch, two different elastic moduli would be required to describe the deformation for triblocks and homopolymers. 

\begin{figure}[t]
\begin{center}
\includegraphics[width=0.5\linewidth]{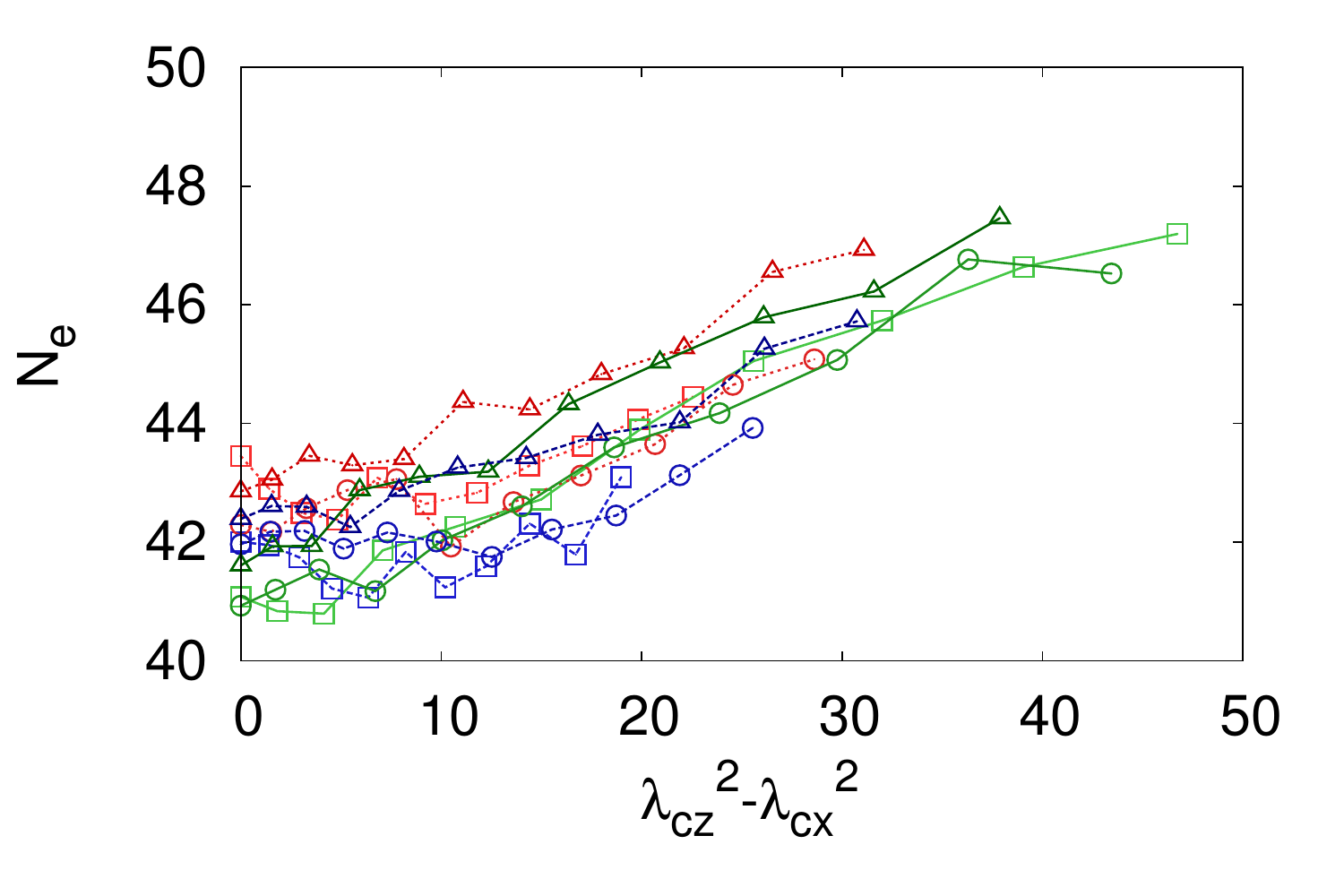}\\
\caption{
\label{ne}
Entanglement length (from Z1 analysis) vs effective stretch. 
Green: triblocks, red: homopolymers, blue: cut chains for $N=300$ ($\square$), $N=500$ ($\circ$), and $N=800$ ($\triangle$).}
\end{center}
\end{figure}

We next consider if entanglement loss could describe the differing stress responses of triblocks and homopolymers. We use the Z1-method \cite{kro05,sha07,hoy09,kar09} to monitor the entanglement length $N_e$ during deformation. This algorithm geometrically minimizes a shortest minimally connected path (SP) for each chain simultaneously. Kinks are entanglement points where chains interact, and their number decreases during the minimization process. The final number of kinks per chain $N_k$ is related to the entanglement length by $N_k=N/N_e$.  
Throughout the deformations we find in Figure \ref{ne} that the entanglement length increases for all chains by less than 10\%. There is no difference in entanglement loss when comparing the homopolymers, triblocks and cut chains other than small variations in initial values $N_e\approx 42$.       

Given this minimal entanglement loss, we now investigate how the polymer network responds to deformation on the length scale of the entanglements. We take the initial set of monomers defined by the kinks of the SP from the Z1 analysis.  The end points of the chains are included as kinks only in the triblock case. We define entanglement vectors $\mathbf{R_k}$ between these monomers and track them during the deformation. 

Typically network models focus on deriving the relationship between entanglements and crosslinks with the applied deformation. However, we can determine this directly through simulation. This allows us to formulate a model for the stress in terms of the entanglement and crosslink stretches. Here, the entanglement length $N_e$ (and therefore number of kinks $N_k$) is approximated as constant at the initial value.    

The \hl{end-to-end} distances $R_k$ and $R_c$ are statistically independent variables. We check this by confirming $\langle R_k\rangle \langle R_c\rangle=\langle R_k R_c\rangle$ throughout the deformations (see \hl{ Figure S1}). The entropies related to the number of possible configurations at a given $R_c$ and $R_k$ are therefore additive, and we can consider the triblock network to be the superposition of two networks on different length scales. The first, consisting of the chain segments between entanglements, is also applicable to homopolymers. Each segment between two entanglement points has $N_e$ beads, with end-to-end distance $R_k$ that is initially $R_k^2(0)=c_\infty N_eb^2$, where $b$ is the FENE bond length \hl{and $c_{\infty}$ is the characteristic ratio.} 
The longer crosslink network enforced by the glassy regions on the triblock chains, has initial end-to-end separation $R_c^2(0)=c_\infty Nb^2$. We do not differentiate between chains that bridge between two glassy regions and those that loop back. \hl{The proportion of bridging chains is $0.85$}, but while the initial end-to-end distances are dramatically different for looping and bridging chains, we find that both their stretches are equal as well the number of kinks found by the Z1-analysis.   
 
For a network of chains of length $N$ crosslinked at their ends, the (purely entropic) free energy density can be written as \cite{arr93}
\begin{equation}
W=G N \left (h \mathcal{L}^{-1}(h) + \textrm{ln}\frac{\mathcal{L}^{-1}(h)}{\textrm{sinh}(\mathcal{L}^{-1}(h))} \right),
\end{equation}
where $G=\nu k_BT$ with the chain density $\nu$, $h$ is the ratio of the chain end-to-end distance to the maximum possible end-to-end separation, and $\mathcal{L}^{-1}$ is the inverse Langevin function. Given that $R_c$ and $R_k$ are statistically independent, the total energy density can be written as the sum of contributions from entanglements and crosslinks. 
We use the 8-chain model to link the end-to-end chain stretch $\lambda_{\textrm{chain}}=h\sqrt{N}$ to the component stretches, $\lambda_{\textrm{chain}}=\sqrt{\frac{1}{3}(\lambda_{x}^2+\lambda_{y}^2+\lambda_{z}^2)}$. We confirm that this relationship holds for both chain stretch $\lambda_c=R_{c}/R_{c}(0)$ and the entanglement stretch  $\lambda_k=R_{k}/R_{k}(0)$ (see \hl{Figure S2}).

\begin{figure}[t]
\begin{center}
\includegraphics[width=0.5\linewidth]{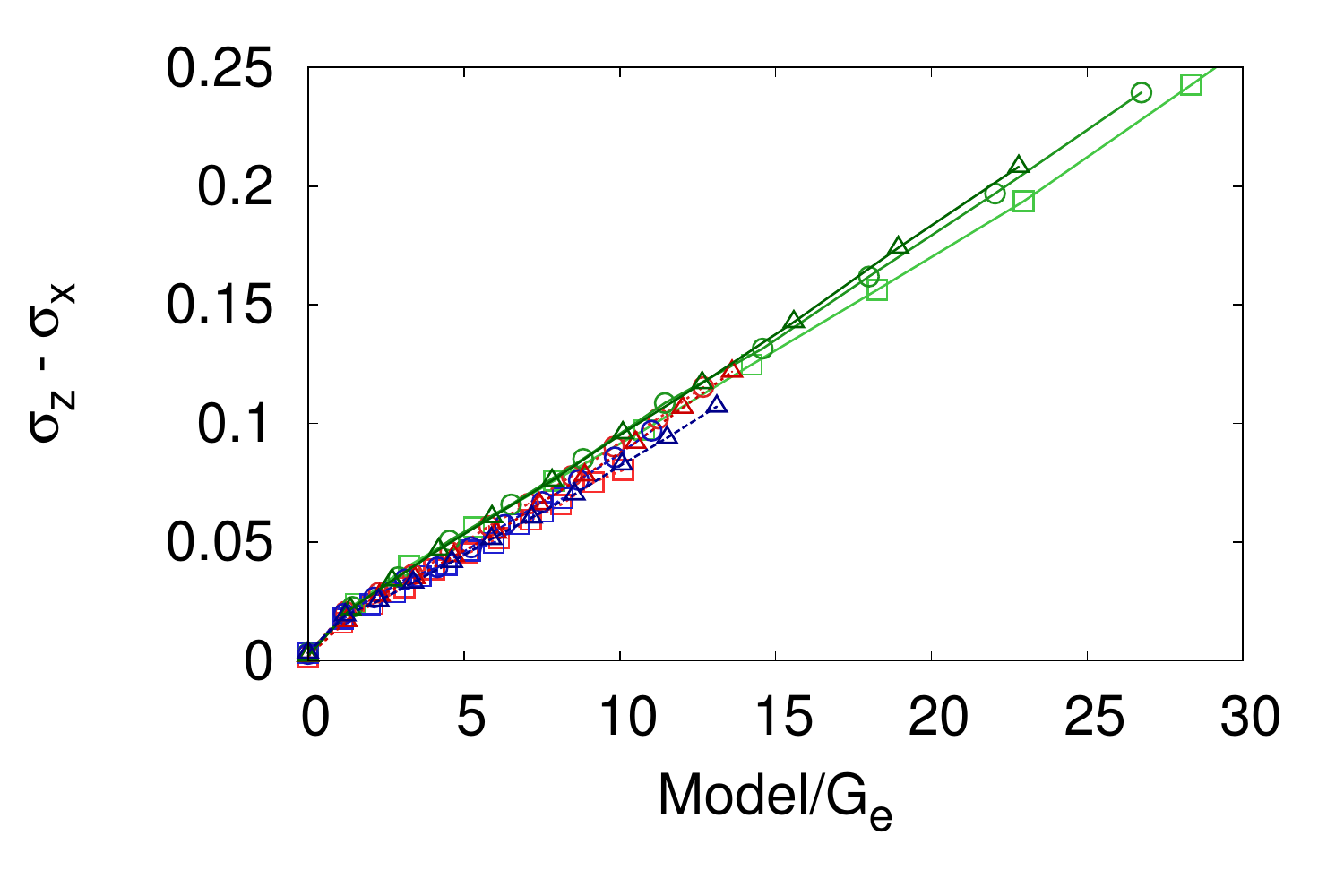}
\caption{\label{le} Simulation stress vs. calculated stress from models for triblocks \hl{(eq.}~\eqref{tri-eq}) and homopolymers/cut chains \hl{(eq.}~\eqref{homcut-eq}).  Green: triblocks, red: homopolymers, blue: cut chains for $N=300$ ($\square$), $N=500$ ($\circ$), and $N=800$ ($\triangle$).}
\end{center}
\end{figure}
The stress contributions can then be found by differentiating the energy density with respect to the deformation of the respective chains,
\begin{equation}
\sigma_i=\lambda_{i}\frac{dW}{d\lambda_{i}} + c =
G \frac{\mathcal{L}^{-1}(h)}{3h}\lambda_{i}^2 + c.
\end{equation}

Entropic network models typically have this form, with contributions due to entanglements and crosslinks, and fit the two elastic moduli $G_e$ and $G_c$ to experimental or simulation data. However, the moduli are related and we can avoid fitting both. In terms of monomer number density $\rho$, 
$G_e=\rho k_BT/N_e$ and $G_c=\rho k_BT/N= G_e/N_k$. In the triblock system, both entanglements and crosslinked ends contribute additively, hence 
\begin{align}
\sigma_{z} -\sigma_{x}&= {G_e}\left[\frac{\mathcal{L}^{-1}(h_k)}{3h_k}(\lambda_{kz}^2-\lambda_{kx}^2) + \frac{1}{N_k}\frac{\mathcal{L}^{-1}(h_c)}{3h_c}(\lambda_{cz}^2 -\lambda_{cx}^2)\right], 
\label{tri-eq}
\end{align}   
where $h_c=R_c/Nb$ and $h_k=R_k/N_eb$. 
\hl{For small $h < 0.4$, the inverse Langevin function can be approximated by $\mathcal{L}^{-1}(h)\approx 3h$. Since the homopolymers and cut chains only display Gaussian hardening,} (se Fig.~\ref{stress}), \hl{and $h_k<0.4$ (see Fig. S3(a)), eq.}(\ref{tri-eq}) \hl{simplifies to the affine network model in terms of the entanglement stretch for these systems}, 
\begin{equation}
\sigma_z-\sigma_x=G_e(\lambda_{kz}^2-\lambda_{kx}^2).
\label{homcut-eq}
\end{equation}
\hl {For the triblocks by contrast, the maximum values of $h_k$ and$h_c$ at the end of the deformations are $0.6$ and $0.55$, so the non-linear regime of the inverse Langevin function is required (for approximants see} ref.~\cite{kro15}).   

In order to test these models, we plot  the measured stresses against the predicted values in Figure \ref{le}. Here, the values for the stretch components, $h_k$, and $h_c$ are input into eqs.~\eqref{tri-eq} and \eqref{homcut-eq} directly from the simulations. For all systems, the data collapses onto a single curve that indicates a nearly perfectly linear relationship. This excellent agreement is strong evidence that we have correctly identified all relevant microscopic chain deformations. The slope of this master curve furthermore has a value of $G_e=0.009$, which agrees well with the expected value of $G_e=\rho k_BT/N_e=0.007$.     

\begin{figure}[t]
\begin{center}
\includegraphics[width=0.5\linewidth]{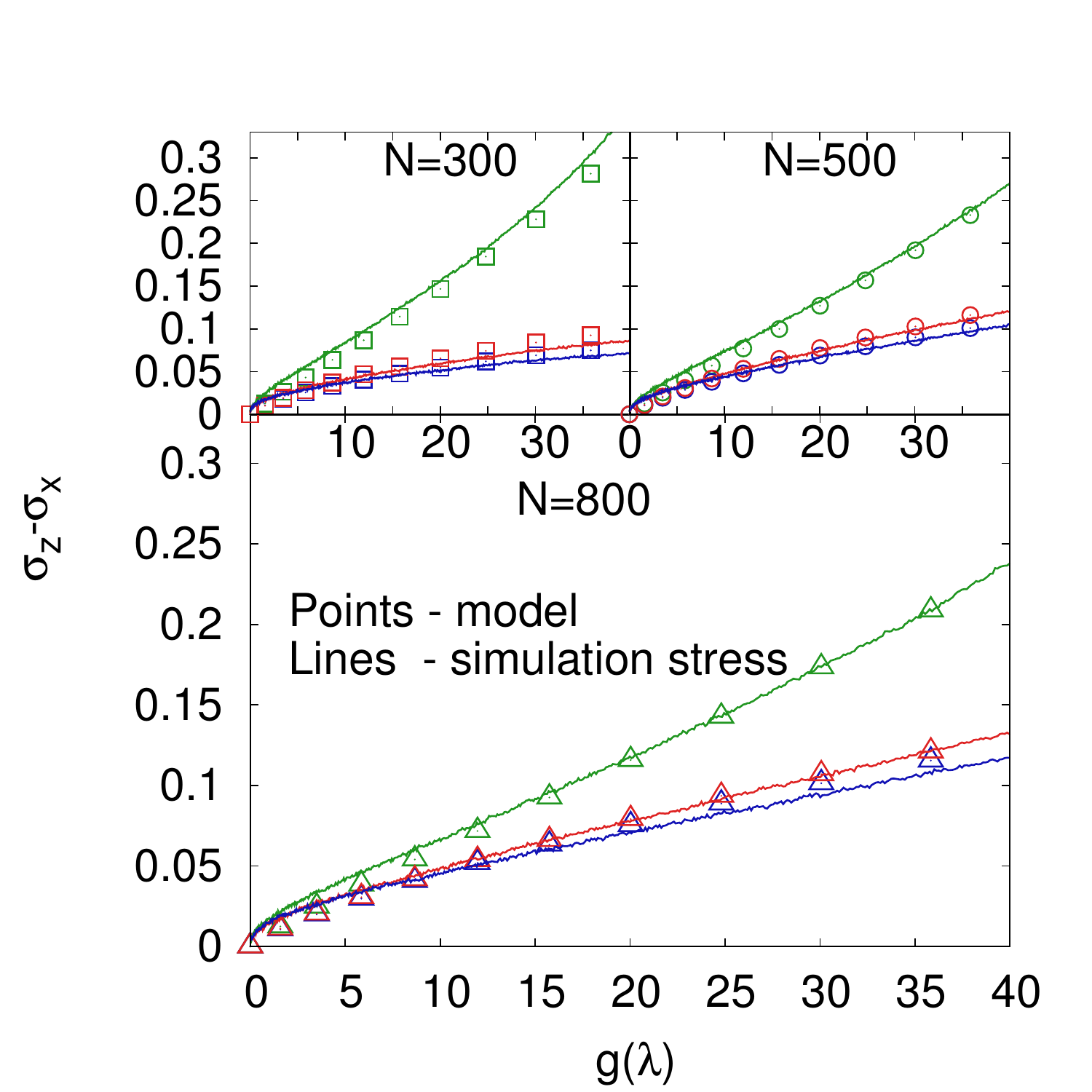}
\caption{\label{fit} Comparison of model (points) to simulation data (lines) in terms of $g(\lambda)=\lambda^2-1/\lambda$. Green: triblocks, blue: homopolymers, red: cut chains for $N=300$ ($\square$), $N=500$ ($\circ$), and $N=800$ ($\triangle$).}
\end{center}
\end{figure}

As a further illustration of the success of our description, we compare in Figure \ref{fit} model and simulated stresses in terms of the global stretch $\lambda$.  Again, we find excellent full quantitative agreement between model prediction and stress-strain curves for all systems and chain lengths studied. Our theory uses only microscopic deformation variables, and establishes an almost fit-free description of the nonlinear mechanical response of nanostructured  polymeric elastomers.


{\bf Acknowledgments}

We thank M.~Kr\"oger for providing us with a copy of the Z1-code. This research was supported by the Natural Sciences and Engineering Research Council of Canada and undertaken thanks in part to funding from the Canada First Research Excellence Fund, Quantum Materials and Future Technologies Program.
\newpage
{\bf Supporting Information:}

The supporting information presents an analysis of the correlations between entanglement and chain stretches, a test of the 8-chain model, and plots of the ratios of the chain end-to-end distance to the maximum possible end-to-end separations.


\providecommand{\latin}[1]{#1}
\providecommand*\mcitethebibliography{\thebibliography}
\csname @ifundefined\endcsname{endmcitethebibliography}
  {\let\endmcitethebibliography\endthebibliography}{}

\end{document}